\documentclass[twocolumn,english,prx, longbibliography]{revtex4-1}
\usepackage[T1]{fontenc}
\usepackage[latin9]{inputenc}
\usepackage{color}
\usepackage{amsbsy}
\usepackage{amssymb}
\usepackage{graphicx}
\usepackage{babel}
\begin{document}
\title{Experimental Demonstration of Exceptional Points of Degeneracy in
Linear Time Periodic Systems and Exceptional Sensitivity}
\author{Hamidreza Kazemi$^{1}$, Mohamed Y. Nada$^{1}$, Alireza Nikzamir$^{1}$,
Franco Maddaleno$^{2}$, Filippo Capolino$^{1}$}
\affiliation{$^{1}$ Department of Electrical Engineering and Computer Science,
University of California, Irvine, CA 92697, USA}
\affiliation{$^{2}$ Department of Electronics and Telecommunications, Politecnico
di Torino, 10129 Torino, Italy}
\begin{abstract}
We present the experimental demonstration of the occurrence of exceptional
points of degeneracy (EPDs) in a single resonator by introducing a
linear time-periodic variation of one of its components, in contrast
to EPDs in parity time (PT)-symmetric systems that require two coupled
resonators with precise values of gain and loss. In the proposed scheme,
only the tuning of the modulation frequency is required that is easily
achieved in electronic systems. The EPD is a point in a system parameters'
space at which two or more eigenstates coalesce, and this leads to
unique properties not occurring at other non-degenerate operating
points. We show theoretically and experimentally the existence of
a second order EPD in a time-varying single resonator. Furthermore,
we measure the sensitivity of the proposed system to a small structural
perturbation and show that the operation of the system at an EPD dramatically
boosts its sensitivity performance to very small perturbations. Also,
we show experimentally how this unique sensitivity induced by an EPD
can be used to devise new exceptionally-sensitive sensors based on
a single resonator by simply applying time modulation.
\end{abstract}
\maketitle

\section{Introduction}

Sensing and data acquisition is an essential part of many medical,
industrial, and automotive applications that require sensing of local
physical, biological or chemical quantities. For instance, pressure
sensors \citep{Chen2010Wireless,Chen2018Generalized}, temperature
sensors \citep{Trung2016all}, humidity sensors \citep{Feng2015Low},
and bio-sensors on the skin or inside the body have gained a lot of
interest in the recent years \citep{Chen2008Microfabricated,Yvanoff2009feasibility,Chen2014Continuous,Corrie2015Blood,Tseng2018Functional,Zhang2019Noninvasive,Kazemi2019Ultra}.
Thus, various low-profile low-cost highly-sensitive electromagnetic
(EM) sensing systems are desirable to achieve continuous and precise
measurement for the mentioned various applications. The operating
nature of the currently used EM resonant sensing systems are mostly
based on the change in the equivalent resistance or capacitance of
the EM sensor by a small quantity $\delta$ (e.g. 1\%), resulting
in changes of measurable quantities such as the resonance frequency
or the quality factor that vary proportionally to $\delta$ (that
is still in the order of 1\%). The scope of this paper is to show
theoretically and experimentally, a new strategy for sensing that
leads to a major sensitivity enhancement based on a physics concept
rather than just an optimization method, which forms a new paradigm
in sensing technology.

In order to enhance the sensitivity of an EM system we exploit the
concept of exceptional point of degeneracy (EPD) at which the observables
are no longer linearly proportional to a system perturbation but rather
have an $m^{th}$ root dependence with $m$ being the order of the
EPD. Such dependence enhances the sensitivity greatly for small perturbations.
For instance, exploiting an EPD of order 2 as in this paper, if we
change a system capacitance by a small quantity $\delta$ (e.g., 1\%)
then the resonance frequency of a resonator operating at an EPD would
change by a quantity proportional to $\sqrt{\delta}$ (e.g., 10\%),
making this fundamental physical aspect very interesting for sensing
very small amounts of substances.

\begin{figure}
\begin{centering}
\includegraphics[width=3in]{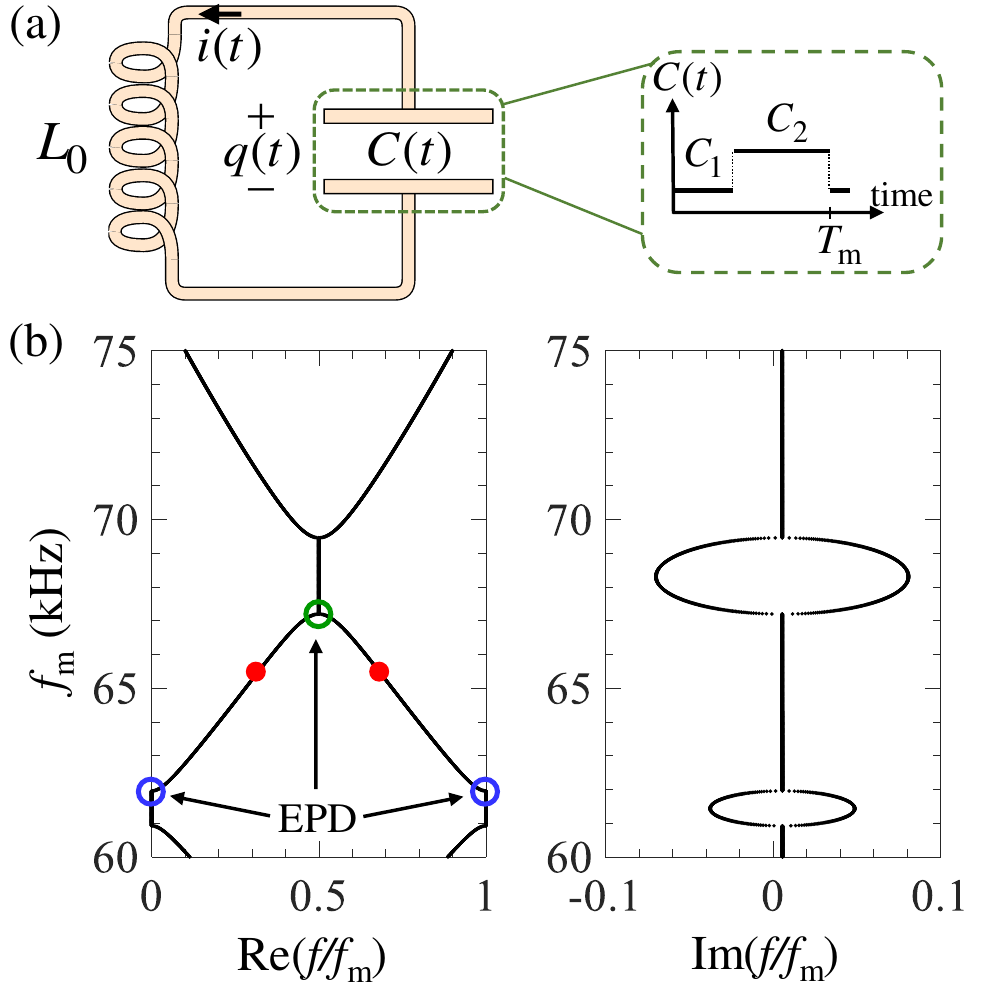}
\par\end{centering}
\caption{\label{fig:Real=000026Imag}(a) Linear time-periodic LC resonator
with a time-varying capacitor. The time-varying capacitance is a periodic
piece-wise constant function as shown in the subset. (b) Dispersion
diagram showing the real and imaginary parts of the eigenfrequencies
of the resonator (i.e., the circuit's resonance frequencies) versus
modulation frequency $f_{\mathrm{m}}$ of the capacitance. Due to
time-periodicity, an EPD resonance at frequency $f_{e0}$ has Fourier
harmonics $f_{e0}+sf_{\mathrm{m}}$, with $s=0,\pm1,\pm2,...$.}
\end{figure}
An EPD of order two is the splitting point (or degenerate point) of
two resonance frequencies and it emerges in systems when two or more
eigenmodes coalesce into a single degenerate eigenmode, in both their
\textit{eigenvalues} and \textit{eigenvectors}. The emergence of EPDs
is associated with unique properties that promote several potential
applications such as enhancing the gain of active systems \citep{Othman2016GiantGain},
lowering the oscillation threshold \citep{Nada2020Exceptionl} or
improving the performance of laser systems \citep{Veysi2018Degenerate,Hodaei2014Parity}
or circuit oscillators \citep{Oshmarin2019New}, enhancing circuits'
sensitivity at radio frequencies \citep{Schindler2011Experimental,Chitsazi2017Experimental,Chen2018Generalized,Sakhdari2018Ultrasensitive,Sakhdari2019Experimental}
or at optical frequencies \citep{Wiersig2014Enhancing,Wiersig2016Sensors,Hodaei2017Enhanced,Chen2017Exceptional},
etc.

EPDs emerge in EM systems using various methods: by introducing gain
and loss in the system based on the concept of parity-time (PT-) symmetry
\citep{Bender1998Real,Stehmann2004Observation,Schindler2011Experimental,Hodaei2014Parity,Nada2018Microwave,Chen2018Generalized,Sakhdari2018Ultrasensitive,Sakhdari2019Experimental,Chitsazi2017Experimental,Wiersig2016Sensors,Wiersig2014Enhancing,Hodaei2017Enhanced,Chen2017Exceptional},
or by introducing periodicity (spatial or temporal periodicity) in
waveguides \citep{Nada2017Theory,Othman2016Giant,Figotin2005Gigantic,Nada2018Various}.
Electronic circuits with EPD based on PT symmetry
have been demonstrated in \citep{Stehmann2004Observation,Schindler2011Experimental}
where the circuit is made of two coupled resonators with loss-gain
symmetry, and only one precise combination of parameters leads to
an EPD. The concept has been further elaborated in \citep{Sakhdari2018Ultrasensitive,Chen2018Generalized}
focusing on the high sensitivity of the EPD circuits introduced in
\citep{Stehmann2004Observation,Schindler2011Experimental} to perturbations.
Note that EPDs realized in PT-symmetric systems require at least two
coupled resonators, and the precise knowledge and symmetry of gain
and loss in the system. In contrast, in this paper we experimentally
demonstrate EPDs that are directly induced via time modulation of
a component in a \textit{single}
resonator \citep{Kazemi2019Exceptional}. An EPD induced by time modulation
in a single resonator is easily tuned by just changing the modulation
frequency of a component: this is a simple and viable strategy to
obtain EPDs since the accurate change of a modulation frequency is
common practice in electronic systems. Moreover,
considering the fact that tuning of the modulation frequency is the
key parameter to get an EPD, this strategy is also immune to tolerances
in the values of commercially available inductors and capacitors.

In this paper we focus on the new scheme to obtain
a second order EPD induced in a linear time periodic (LTP) system
as introduced in \citep{Kazemi2019Exceptional}. Here EPDs are obtained
by applying the time-periodic modulation to a system parameter (i.e.,
the capacitor) in a \textit{single}
resonator, and we provide an experimental demonstration of the existence
of the LTP-induced EPD. Moreover, we show theoretically and experimentally
how the resonance frequencies of a single EPD resonator are strongly
perturbed by a tiny perturbation of its capacitor, and explore possible
sensing applications of such phenomenon. We show that the system's
resonance frequency shift generated by a perturbed EPD follows the
Puiseux fractional power expansion series \citep{Welters2011Explicit},
i.e., if $\delta$ is a perturbation to a second order EPD system,
two resonances arise shifted by a quantity proportional to $\sqrt{\delta}$
from the EP degenerate resonance frequency. On the other hand, perturbed
systems not operating at an EPD exhibit a frequency shift proportional
to $\delta$, that is much smaller than $\sqrt{\delta}$ when $\left|\delta\right|\ll1$.
The theoretical predictions are in excellent agreement with the experimental
demonstration, showing that tiny system perturbations can be detected
by easily measurable resonance frequency shifts, even in the presence
of electronic and thermal noise.

\section{Formulation of an LTP-induced EPD in a Single Resonator\label{sec:LPT_EPD_Theory}}

We investigate resonances and their degeneracy in a linear time-periodic
(LTP) LC resonator as shown in Fig. \ref{fig:Real=000026Imag}(a)
where the time-varying capacitance is shown in the figure subset.
This single-resonator circuit is supporting an EPD induced by the
time-periodic variation. We consider a piece-wise constant time-varying
capacitance $C(t)$ with period $T_{\mathrm{m}}$; we have chosen
the piece-wise function to make the theoretical analysis easier, yet
the presented analysis is valid for any periodic function. A thorough
theoretical study of this type of temporally induced EPDs has been
presented in \citep{Kazemi2019Exceptional}, here we focus on the
energy transfer formulation of these EPDs, and we show the first practical
implementation of the EPDs induced in LTP systems.

The state vector $\mathbf{\boldsymbol{\Psi}}(t)$ describing the system
in Fig. \ref{fig:Real=000026Imag} is two-dimensional (see Ref. \citep{Kazemi2019Exceptional}
for $N$-dimensional), i.e., $\mathbf{\boldsymbol{\Psi}}(t)=[q(t),i(t)]^{\mathrm{T}}$,
where $\mathrm{T}$ denotes the transpose operator, $q(t)$ and $i(t)$
are the capacitor charge and inductor current, respectively. The temporal
evolution of the state vector obeys the 2-dimensional first-order
differential equation

\begin{equation}
\frac{\mathrm{d}\mathbf{\boldsymbol{\Psi}}(t)}{\mathrm{d}t}=\mathbf{\underline{M}}(t)\mathbf{\boldsymbol{\Psi}}(t),\label{eq:time_evolu}
\end{equation}
where $\mathbf{\underline{M}}(t)$ is the $2\times2$ system matrix.
The 2-dimensional state vector $\boldsymbol{\Psi}(t)$ is derived
at any time $t=nT_{\mathrm{m}}+\chi$ with $n$ being an integer and
$0<\chi<T_{\mathrm{m}}$ as

\begin{equation}
\boldsymbol{\Psi}(t)=\underline{\boldsymbol{\Phi}}(\chi,0)\left[\underline{\boldsymbol{\Phi}}(T_{\mathrm{m}},0)\right]^{n}\boldsymbol{\Psi}(0),\label{eq:Psi(t)}
\end{equation}
where $\underline{\boldsymbol{\Phi}}(t_{2},t_{1})$ is the $2\times2$
state transition matrix \citep{Kazemi2019Exceptional} that translates
the state vector from the time instant $t_{1}$ to $t_{2}$. The state
transition matrix is employed to represent the time evolution of the
state vector, hence we formulate the eigenvalue problem as \citep{Kazemi2019Exceptional}

\begin{equation}
(\underline{\boldsymbol{\Phi}}-\lambda\underline{\mathbf{I}})\boldsymbol{\Psi}(t)=0.\label{eq:eigen value prob}
\end{equation}

The two eigenvalues are $\lambda_{p}=\exp(\mathrm{j}2\pi f_{p}T_{\mathrm{m}})$,
$p=1,2$, where $f_{p}$ are the system resonant frequencies, with
all their Fourier harmonics $f_{p}+sf_{\mathrm{m}}$, where $s$ is
an integer and $f_{\mathrm{m}}=1/T_{\mathrm{m}}$ is the modulation
frequency. Fig. \ref{fig:Real=000026Imag}(b) shows the dispersion
diagram of the resonant frequencies varying modulation frequency $f_{\mathrm{m}}$,
restricting the plot to frequencies in the range $0<f/f_{\mathrm{m}}<1$,
which is called here as the fundamental Brillouin zone (BZ), adopting
the language used in space-periodic structures. The two red circles
in Fig. \ref{fig:Real=000026Imag}(b) corresponds to two general distinct
resonant frequencies at $f$ and $-f+f_{\mathrm{m}}$ with positive
real part. An EPD occurs when these two resonance frequencies coalesce
at a given modulation frequency. At an EPD the transition matrix $\underline{\boldsymbol{\Phi}}$
is non-diagonalizable with a degenerate eigenvalue $\lambda_{e}$
(with corresponding eigenfrequency $\omega_{e}$) because the two
eigenvalues and two eigenvectors coalesce. Two possibilities may occur
because of the nature of the problem (time periodicity, because there
are only two possible eigenmodes, and neglecting losses for a moment):
the degenerate eigenvalue is either i) $\lambda_{e}=-1,$ corresponding
to an eigenfrequency of $f_{e0}$ =$f_{\mathrm{m}}/2$ and its Fourier
harmonics $f_{es}=f_{e0}+sf_{\mathrm{m}}$ shown with the green circles
in Fig. \ref{fig:Real=000026Imag}(b), or ii) $\lambda_{e}=1$, corresponding
to $f_{e0}=0$ and its Fourier harmonics $f_{es}=f_{e0}+sf_{\mathrm{m}}$
shown with a blue circle in Fig. \ref{fig:Real=000026Imag}(b). Therefore,
in general an EPD resonance is characterized by the fundamental frequency
$f_{e0}$ and all harmonics at $f_{es}=f_{e0}+sf_{\mathrm{m}}$ where
$s=0,\pm1,\pm2,...$. Moreover, at an EPD the state transition matrix
$\underline{\boldsymbol{\Phi}}$ is similar to a Jordan-Block matrix
of second order, hence it has a single eigenvector, i.e., the geometrical
multiplicity of the eigenvalue $\lambda_{e}$ is equal to 1 while
its algebraic multiplicity is equal to 2. Considering the circuit
parameters given in the next section including inductor small series
resistance, an EPD occurs at $f_{\mathrm{m}}=62.7\,\mathrm{kHz}$,
leading to a degenerate resonance frequency of $f_{e0}/f_{\mathrm{m}}=\mathrm{j}0.0041$
which corresponds to blue circle in Fig. \ref{fig:Real=000026Imag}(b).
Another EPD occurs at $f_{\mathrm{m}}=67.7\,\mathrm{kHz}$, leading
to a degenerate resonance frequency of $f_{e0}/f_{\mathrm{m}}=0.5+\mathrm{j}0.0038$
which corresponds to green circle in Fig. \ref{fig:Real=000026Imag}(b).

When $\lambda_{e}=-1$ and hence $f_{e0}$ =$f_{\mathrm{m}}/2$, i.e.,
for EPDs at the center of the BZ, the state transition matrix $\underline{\boldsymbol{\Phi}}$
has a trace of $-2$ \citep{Kazemi2019Exceptional}, so that we express
$\left[\underline{\boldsymbol{\Phi}}(T_{\mathrm{m}},0)\right]^{n}$as
\citep{Richards1983Analysis}

\begin{equation}
\left[\underline{\boldsymbol{\Phi}}(T_{\mathrm{m}},0)\right]^{n}=\left(-1\right)^{n+1}[n\boldsymbol{\underline{\Phi}}(T_{\mathrm{m}},0)+(n+1)\mathbf{\underline{I}}],\label{eq:Phi^n}
\end{equation}
where $\mathbf{\underline{I}}$ is the $2\times2$ identity matrix.
Thus, we reformulate (\ref{eq:Psi(t)}) using (\ref{eq:Phi^n}) as

\begin{equation}
\boldsymbol{\Psi}(t)=\left(-1\right)^{n+1}[n\boldsymbol{\underline{\Phi}}(T_{\mathrm{m}},0)+(n+1)\mathbf{\underline{I}}]\boldsymbol{\underline{\Phi}}(\chi,0)\boldsymbol{\Psi}(0).\label{eq:Final Psi(t)}
\end{equation}
\begin{figure}
\begin{centering}
\includegraphics[width=3in]{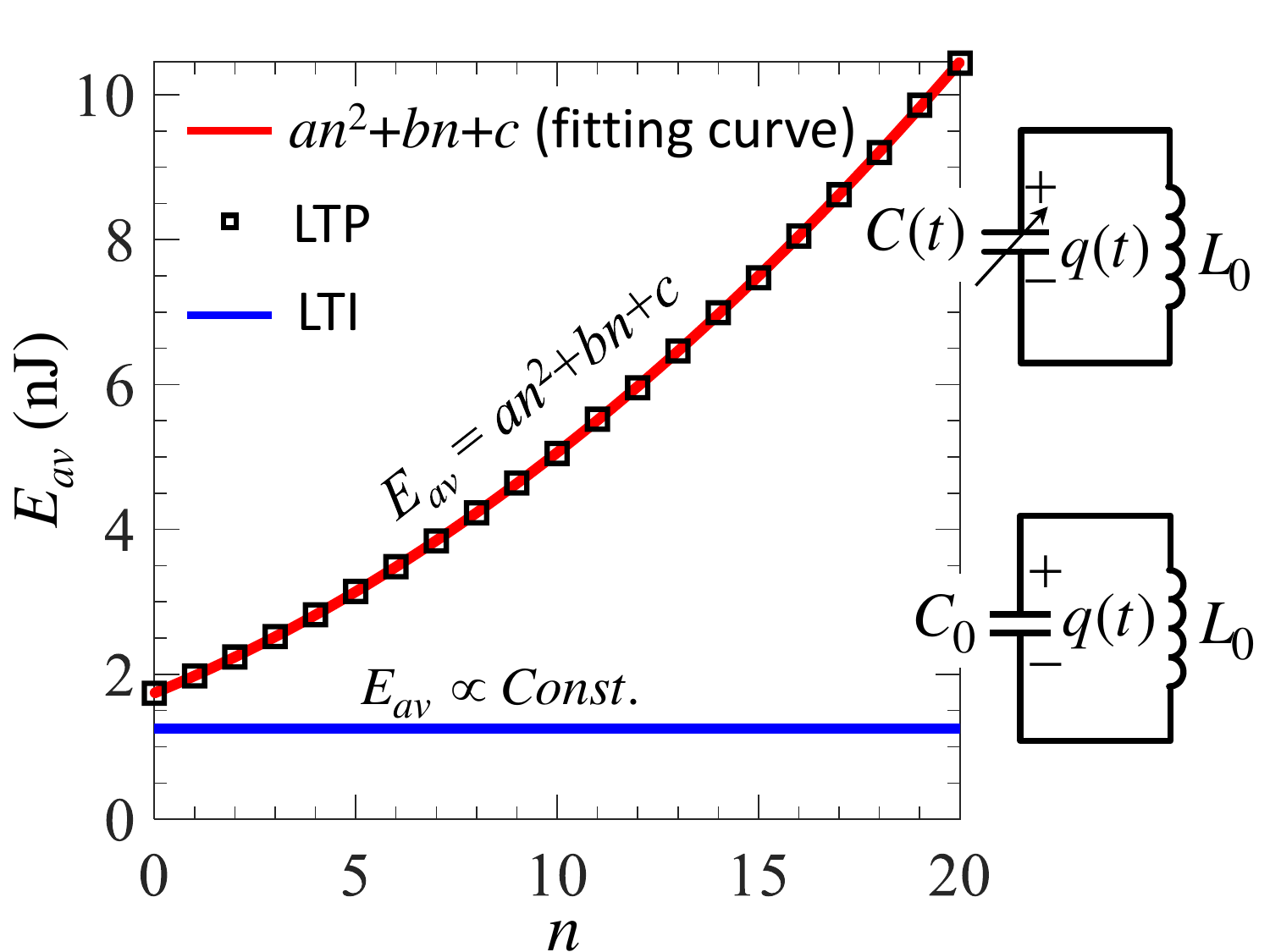}
\par\end{centering}
\caption{\label{fig:Engergy_Trans}Comparison between the time-average energy
stored in a lossless, linear time-invariant LC resonator (solid blue
line) and the time-average energy stored in a linear time-periodic
LC resonator operating at an EPD located at the edge of BZ (black
square symbols) where $n$ is an integer representing the number of
elapsed modulations periods. In the latter case the time-average energy
grows with time, fitted by a second order polynomial curve (red solid
line). The fitting coefficients are set as $a=0.01$, $b=0.23$, and
$c=1.74$.}
\end{figure}
Similarly, for EPDs at the edge of the BZ, i.e., when $\lambda_{e}=1$
and $f_{e0}=0$, the transition matrix $\underline{\boldsymbol{\Phi}}$
has a trace equal to $2$, and \citep{Richards1983Analysis}

\begin{equation}
\left[\underline{\boldsymbol{\Phi}}(T_{\mathrm{m}},0)\right]^{n}=n\boldsymbol{\underline{\Phi}}(T_{\mathrm{m}},0)-(n-1)\mathbf{\underline{I}},
\end{equation}
hence

\begin{equation}
\boldsymbol{\Psi}(t)=[n\boldsymbol{\underline{\Phi}}(T_{\mathrm{m}},0)-(n-1)\mathbf{\underline{I}}]\underline{\boldsymbol{\Phi}}(\chi,0)\boldsymbol{\Psi}(0).\label{eq:Psi_Edge}
\end{equation}

Because of the multiplication of the time-period step \textit{$n$},
we conclude from Eqs. (\ref{eq:Final Psi(t)}) and (\ref{eq:Psi_Edge})
that when the system is at the second order time-periodic induced
EPD, the state vector grows linearly with time. This linear growth
is expected and it is one of the unique characteristics associated
with EPDs. This algebraic growth is analogous to the spatial growth
of the state vector associated with the space-periodic EPDs \citep{Othman2016GiantGain,Figotin2005Gigantic}.

The time-periodic LC tank considered in this section is not ``isolated''.
In such a system, the time-varying capacitor is in continuous interaction
with the source of the time variation that is exerting work. This
interaction leads to a net energy transfer into or out of the LC tank;
at some operating modulation frequencies the system simply loses energy
to the time-variation source, while at other operating modulation
frequencies the LC tank receives energy from the source of time variation.
This behavior is in contrast to the behavior of a time-invariant lossless
LC tank where the initial energy in the system is conserved and the
net energy gain or loss is zero. The average transferred energy into
or out of the time-periodic LC tank can be calculated using the time-domain
solution of the two-dimensional first-order differential equation
(\ref{eq:Psi_Edge}). Figure \ref{fig:Engergy_Trans} shows the calculated
time-average energy transferred into a linear time-invariant (LTI)
lossless LC tank (solid blue line) and into an LTP one operating at
a second order EPD (black square symbols), where $n=0$ shows the
average energy of the systems within the first period. The capacitor
in both systems is initially charged with an initial voltage of $V_{C}(0^{-})=-50\,\mathrm{m}\mathrm{V}$.
In the LTP system, the modulation frequency is adjusted to $f_{\mathrm{m}}=62.7\,\mathrm{kHz}$,
so that it operates at the EPD denoted by the blue circles in Fig.
\ref{fig:Real=000026Imag}(b). Note that the system is periodic, so
that for an eigenfrequency $f_{p}$, there are also all the Fourier
harmonics with frequencies $f_{p}+sf_{\mathrm{m}}$, where $s$ is
an integer \citep{Kazemi2019Exceptional}. It is clear that the total
energy in a lossless time-invariant LC resonator is constant over
time while the average energy in the time-periodic LC resonator is
growing at the EPD. This energy growth is quadratic in time since
the state vector (i.e., capacitor charge and inductor current) of
a periodically time-variant LC resonator experiencing an EPD grows
linearly with time, as shown in Eqs. (\ref{eq:Final Psi(t)}) and
(\ref{eq:Psi_Edge}). Indeed, the solid red curve in Fig. \ref{fig:Engergy_Trans}
shows a second order polynomial curve fitted to the LTP LC resonator
energy where the fitting coefficients are given in the figure caption.
One may note from the figure that the average energy of the LTI and
LTP systems are not equal at $n=0$ which might seem counter intuitive.
In fact, at $n=0$ we show the average energy of the systems within
the first time-period which is higher for the LTP system due to the
energy transfer within that first period.

Note that the time-varying capacitance in our proposed
scheme has some resemblance to the concept of parametric amplification
\citep{Arthur1960Parametric,Cassedy1963Dispersion,R.C.Honey1960Wide,Lee1997Degenerate,Yamamoto2008Flux}.
However, in our single resonator proposed scheme, we use time-periodicity
of a system parameter to achieve an EPD \citep{Kazemi2019Exceptional},
and show high sensitivity of the degenerate
resonance to system perturbations. In contrast, parametric amplifiers
use time variation of a component as a non-conservative process to
inject energy and generate amplification (which is not the case in
our circuit), and generally possess low sensitivity to perturbations.

\section{Experimental Demonstration of an EPD and its Sensitivity to Perturbations\label{sec:Experimental-demonstration}}

\begin{figure}
\begin{centering}
\includegraphics[width=3in]{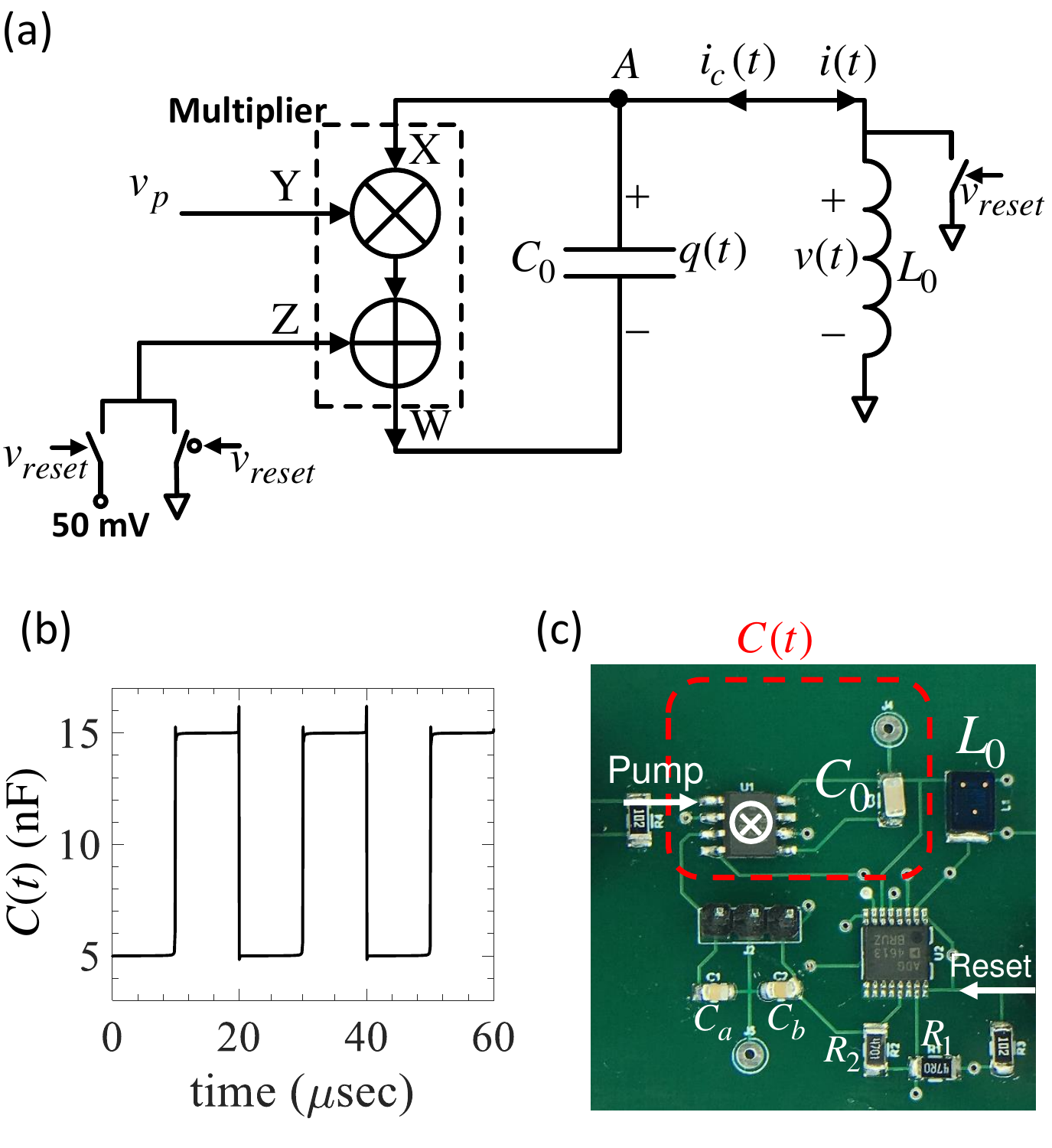}
\par\end{centering}
\caption{\label{fig:CKT_schem_actual}(a) Schematic of the LTP-varying LC resonator
using the periodic pump voltage $v_{p}(t)$ and a multiplier. The
reset switches are used in the implementation to avoid saturation
and to add the initial voltage at the beginning of each ``run''
time. (b) Time domain simulations of the time-varying synthetic capacitance
with period $T_{\mathrm{m}}$ seen from Node $A$ with respect to
the ground. (c) Assembled circuit where the red dashed square shows
the synthetic time-varying capacitor using the pump voltage and a
multiplier. The circuit also consists of the inductor $L_{0}$, the
reset circuit with switches, the regulating capacitors $C_{a}$ and
$C_{b}$, and the circuitry to produce the initial voltage (see Supplement
Material).}
\end{figure}

In this section we verify \textit{experimentally} the key properties
inferred from the degeneracy of resonances shown in the dispersion
diagram in Fig. \ref{fig:Real=000026Imag} and we observe some interesting
physical properties by providing an initial charge to the capacitor
and measure the triggered time domain natural response. The time varying
LC tank with the time-periodic capacitor is implemented based on the
scheme shown in Fig. \ref{fig:CKT_schem_actual}(a). The time variation
is carried out using a time-varying \textit{pump} voltage $v_{p}(t)$
and a multiplier. Hence the voltage applied to the capacitor $C_{0}$
is equal to $v_{c}(t)=v(t)[1-v_{p}(t)/V_{0}]$, where $v(t)$ is the
voltage of Node A with respect to the ground, and the term $V_{0}$
is a constant coefficient of the multiplier that is used to normalize
its output voltage. Here we are interested in having an LC resonator
with a time varying capacitor $C(t)$ seen by the current $i_{c}(t)$
exiting the inductor and by the voltage $v(t)$ at Node $A$, hence
satisfying the two equations $q(t)=C(t)v(t)$ and $\mathrm{d}i_{c}/\mathrm{d}t=-v(t)/L_{0}=-q(t)/(L_{0}C(t))$.
This leads to the definition of the time varying capacitance $C(t)\triangleq C_{0}(1-v_{p}(t)/V_{0})$
and to a LTP LC circuit described by Eq. (\ref{eq:time_evolu}) (see
Supplement Material), which in turns leads to the time domain dynamics
exhibiting EPDs as described in \citep{Kazemi2019Exceptional}.

In such a LTP LC circuit, the time variation behavior of the capacitance
is dictated by the variation of the pump voltage $v_{p}(t)$, therefore
to design the time-varying capacitance shown in Fig. \ref{fig:Real=000026Imag}(a)
we apply a two level piece-wise constant pump voltage to the multiplier.
The values of $C_{1}$ and $C_{2}$ are adjusted by properly choosing
the voltages of the piece-wise constant pump $v_{p}(t)$ as discussed
in the Supplement Material. We aim at designing the time-varying capacitor
$C(t)$ with the values of $C_{1}=5\,\mathrm{n\mathrm{F}}$ and $C_{2}=15\,\mathrm{n\mathrm{F}}$.
Hence, the parameters of the circuit are set as $C_{0}=10\,\mathrm{n\mathrm{F}}$,
the two levels of the piece-wise constant time varying pump voltage
as $v_{p}/V_{0}=\pm0.5$, and the period of the pump voltage as $T_{\mathrm{m}}=20\,\mathrm{\mu s}$
with $50\%$ duty cycle. We verify the operation of this scheme using
the finite difference time domain (FDTD) simulation implemented in
Keysight ADS, where a constant capacitor $C_{0}$ is connected to
the voltage multiplier. The time-varying capacitance, calculated as
the ratio of the current passing through the capacitor $C_{0}$ to
the time-derivative of the voltage at Node $A$, i.e., $i_{c}(t)/[\mathrm{d}v/\mathrm{d}t]$,
is shown in Fig. \ref{fig:CKT_schem_actual}(b). It is worth mentioning
that in such a scheme, the \emph{high} level of the pump voltage $v_{p}$
controls the value of the capacitance $C_{1}$ and the \emph{low}
level controls the value of the capacitance $C_{2}$.

Figure \ref{fig:CKT_schem_actual}(c) illustrates the assembled circuit
where the red dashed rectangle shows the implemented synthetic time-periodic
capacitor $C(t)$. In this fabricated circuit we use a four-quadrant
voltage output analog multiplier , a high stability and precision
ceramic capacitor $C_{0}=10\,\mathrm{n\mathrm{F}}$, and an inductor
$L_{0}=33\,\mu\mathrm{H}$ with low DC resistance $R_{DC}=108\,\mathrm{m\Omega}$,
as specified in the Supplement Material. As shown in Sec. \ref{sec:LPT_EPD_Theory},
we expect the capacitor voltage of the time-periodic LC tank to grow
linearly in time when operating at an EPD; however, in practice it
will saturate to the maximum output voltage of the multiplier. Therefore,
to avoid voltage saturation, we have implemented a reset mechanism
to reset the resonator circuit, where the reset signal is a digital
clock with $20\%$ duty cycle (i.e., $v_{reset}=2\,\mathrm{V}$ for
20\% of its period and $v_{reset}=0\,\mathrm{V}$ otherwise) that
allows the resonator circuit to run for the duration of the low voltage
$v_{reset}=0\,\mathrm{V}$. During the reset time, the reset signal
is high $v_{reset}=2\,\mathrm{V}$, and the resonator circuit is at
pause. At the end of this time interval the capacitor is charged again
with the initial voltage of $V_{C}(0^{-})=-50\,\mathrm{m}\mathrm{V}$
for the start of the next working cycle as detailed in the Supplement
Material. The circuit is provided with $\pm5\,\mathrm{V}$ DC voltage
using a Keysight E3631A DC voltage supply. We use two Keysight 33250A
function generators, one to generate a two level piece-wise constant
signal with levels of $\pm0.525\,\mathrm{V}$, duty cycle of $50\%$
and variable modulation frequency $f_{\mathrm{m}}$ as pump voltage
$v_{p}(t)$ to generate the time-periodic capacitance $C(t)$. The
other function generator provides the resonator's reset signal, a
two level piece-wise constant signal with levels of $2\,\mathrm{V}$
and $0\,\mathrm{V}$, with duty cycle of 20\% and frequency of $1.1\,\mathrm{k}\mathrm{Hz}$
(much lower than $f_{\mathrm{m}}$).

\subsection{Dispersion diagram and time domain response}

\begin{figure}
\begin{centering}
\includegraphics[width=3.5in]{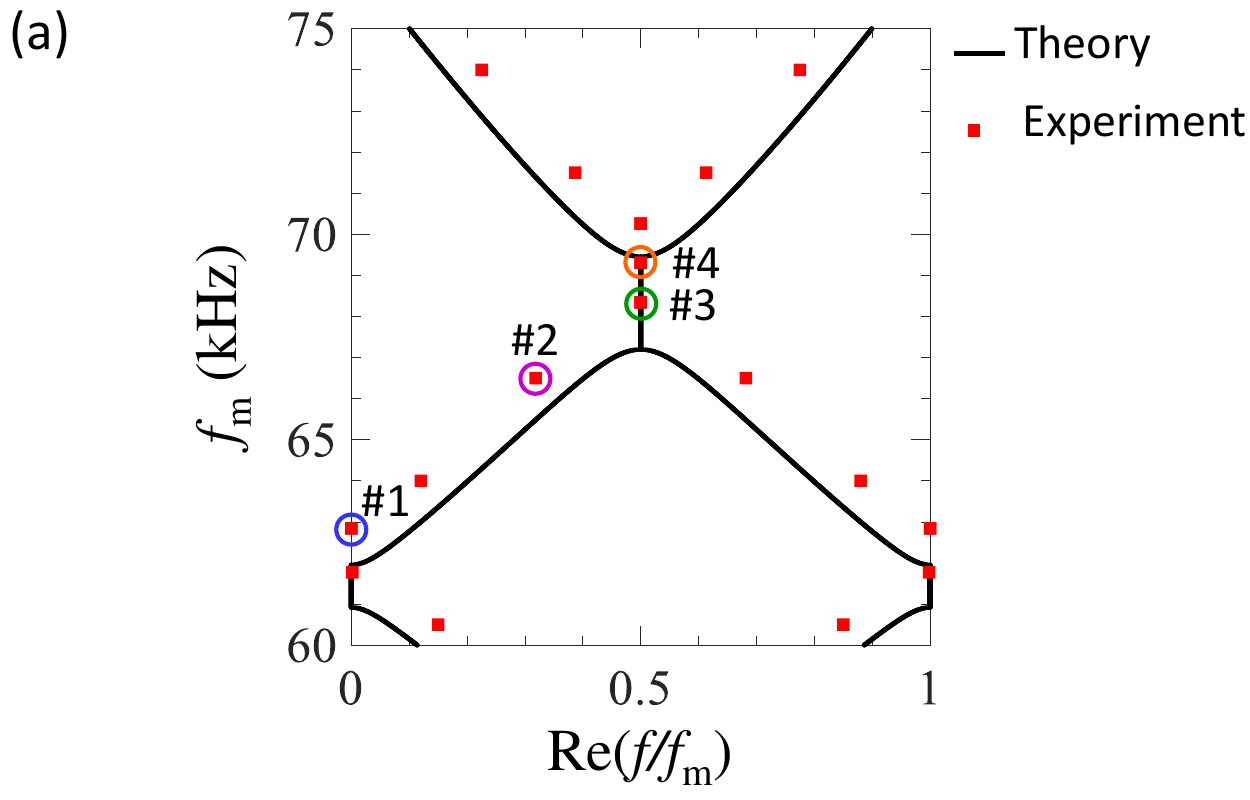}
\par\end{centering}
\begin{centering}
\includegraphics[width=3.5in]{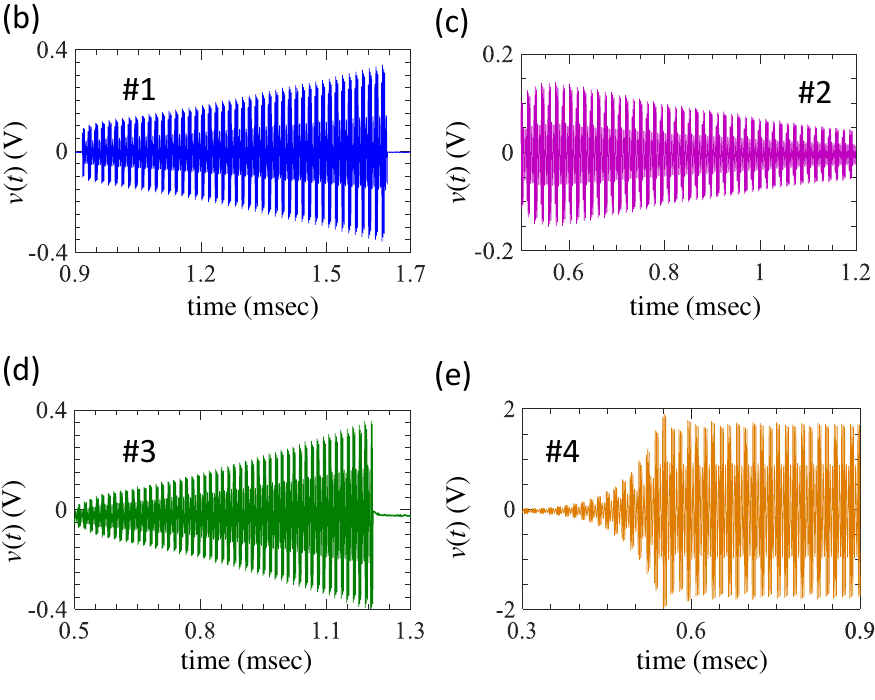}
\par\end{centering}
\caption{\label{fig:exprmnt_results}(a) Theoretical (solid lines) and experimental
(square red symbols) results for the complex dispersion diagram showing
the circuit's resonance frequencies as a function of modulation frequency
$f_{\mathrm{m}}$. (b-e) Time domain experimental results of the capacitor
voltage at various modulation frequencies denoted by \#1-4 in the
dispersion diagram: (b) Exceptional point \#1 denoted by the blue
circle located at the edge of the BZ where the modulation frequency
is $f_{\mathrm{m}}=62.8\,\mathrm{kHz}$. (c) Real-frequency resonance
indicated by \#2, denoted by the magenta circle, where the modulation
frequency $f_{\mathrm{m}}=66.5\,\mathrm{kHz}$. (d) Exceptional point
\#3, denoted by the green circle, located at the center of the BZ
where the modulation frequency is $f_{\mathrm{m}}=68.3\,\mathrm{kHz}$.
(e) Unstable regime, point \#4, denoted by an orange circle, represents
two complex resonance frequencies, with opposite imaginary parts and
real part equal to $f_{\mathrm{m}}/2$, where the modulation frequency
is set to $f_{\mathrm{m}}=71.5\,\mathrm{kHz}.$ Note that due to time-periodicity,
time domain signals have all Fourier harmonics $f+sf_{\mathrm{m}}$,
with $s=0,\pm1,\pm2,...$.}
\end{figure}

Figure \ref{fig:exprmnt_results}(a) presents the dispersion diagram
as a function of the capacitance's modulation frequency $f_{\mathrm{m}}$,
which is experimentally varied by adjusting the frequency of the pump
voltage $v_{p}(t)$. The solid curve denotes the theoretical dispersion
diagram whereas red square symbols represent the experimental results.
The experimental results are obtained by calculating the resonance
frequency of the circuit's response for different modulation frequencies
using Fourier transform of the time domain signal triggered by the
initial voltage $V_{C}(0^{-})$ at each working cycle, where we used
a Keysight DSO7104A digital oscilloscope to capture the time domain
output signal. A good agreement is observed between the theoretical
and experimental results, however, there is a slight frequency shift
between the theoretical and experimental dispersion diagrams which
is due to parasitic reactances, components' tolerances and nonidealities
in the fabricated circuit. Note that in Figure \ref{fig:exprmnt_results}(a)
we show only solutions in the first Brillouin zone defined here as
$\mathrm{Re}(f)\in(0,f_{\mathrm{m}}).$ Since the system is time periodic,
every mode is composed of an infinite number of harmonics with frequencies
$f+sf_{\mathrm{m}}$, where $s$ is an integer. One can observe from
the dispersion diagram that the time-periodic LC resonator operates
at three different regimes depending on the modulation frequency.
In the following we describe the three possible regimes of operation.

i) \textit{Real resonance}s: This is a regime where the system has
two purely real oscillating frequencies (though in practice there
is a small imaginary part due to the finite quality factor of the
components). Point \#2 with the magenta circle in Fig. \ref{fig:exprmnt_results}(a)
illustrates a mode part of this regime, with real resonance frequencies
$f$, and all its harmonics $f+sf_{\mathrm{m}}$, where $s$ is an
integer. The signal has also the resonance frequency at $-f+f_{\mathrm{m}}$,
and hence also the harmonics at $-f+sf_{\mathrm{m}}$. This means
that two resonance modes with frequencies $f$ and $-f+f_{\mathrm{m}}$
are allowed. Fig. \ref{fig:exprmnt_results}(c) shows the corresponding
time domain signal at $f_{\mathrm{m}}=66.5\,\mathrm{kHz}$ which corresponds
to two almost-real resonance frequencies and their harmonics. As mentioned,
the observed small exponential decay of the response is due to the
finite quality factor of the components. In an ideal lossless system
frequencies would be purely real.

ii) \textit{Unstable condition}: This is a regime where the system
has two complex resonance frequencies with imaginary parts of opposite
signs (point \#4 with the orange circle in Fig.\ref{fig:exprmnt_results}(a)).
According to the time convention $\mathrm{e}^{\mathrm{j}\omega t}$,
the state vector of the system corresponding to a complex resonance
frequency with an imaginary part of negative sign shows an exponential
growth, while the state vector corresponding to the resonance with
an imaginary part with positive sign exhibits an exponential decay,
hence it shows an unstable system. The exponentially growing behavior
would be the dominant one and it is the one seen in the time domain
response in Fig. \ref{fig:exprmnt_results}(e) that would eventually
saturate, if we did not include a reset circuit. In this case the
modulation frequency of the pump voltage is set to $f_{\mathrm{m}}=71.5\,\mathrm{kHz}$.
iii) \textit{Exceptional points of degeneracy}: An EPD is the point
that separates the two previous regimes, where two frequency branches
of the dispersion diagram (describing two independent resonance solutions)
coalesce. Indeed at the EPD the two resonant modes of the system coalesce
and, as discussed in the previous section, the state vector shows
a linear growth with time yet the resonance frequencies are real (when
neglecting the small positive imaginary part of the resonance frequency
due to finite quality factor of the components). From the theoretical
analysis and from the experimental results in Fig. \ref{fig:exprmnt_results}(a)
one may note two types of EPDs exist in a linear time-periodic system
\citep{Kazemi2019Exceptional}. EPDs that exist at the center of the
BZ, i.e., at $f_{e0}=0.5f_{\mathrm{m}}$, with Fourier harmonics located
at $f_{es}=\left(0.5+s\right)f_{\mathrm{m}}$, where the integer $s$
denotes the harmonic number. An example of such a type of EPDs is
observed at $f_{\mathrm{m}}=68.3\,\mathrm{kHz}$ and is denoted by
point \#3 with the green circle in Fig. \ref{fig:exprmnt_results}(a).
The measured time domain behavior of the circuit at this modulation
frequency is shown in Fig. \ref{fig:exprmnt_results}(d) where we
clearly see the linear growth of the capacitor voltage. It grows until
it reaches saturation or till the system is reset as described in
the previous section. The other type of EPDs are those that exist
at the edge of the BZ, i.e., at $f_{e0}=0$, with Fourier harmonics
located at $f_{es}=sf_{\mathrm{m}}$. An example of this type of EPDs
is denoted by point \#1 with the blue circle at $f_{\mathrm{m}}=62.8\,\mathrm{kHz}$
in Fig. \ref{fig:exprmnt_results}(a). The measured time domain behavior
of the circuit at such an EPD is depicted in Fig. \ref{fig:exprmnt_results}(b).
Note that the oscillation of the time domain signal for an EPD at
the edge of the BZ is due to the harmonics located at $sf_{\mathrm{m}}$.

One may observe that a standard ``critically damped'' LTI RLC circuit
with two coinciding resonance frequencies is also an exceptional point,
however, that point is characterized by two resonance frequencies
with vanishing real part; hence, it is a different condition from
what we describe in this paper.

\subsection{High sensitivity to perturbations}

\begin{figure}
\begin{centering}
\includegraphics[width=3.2in]{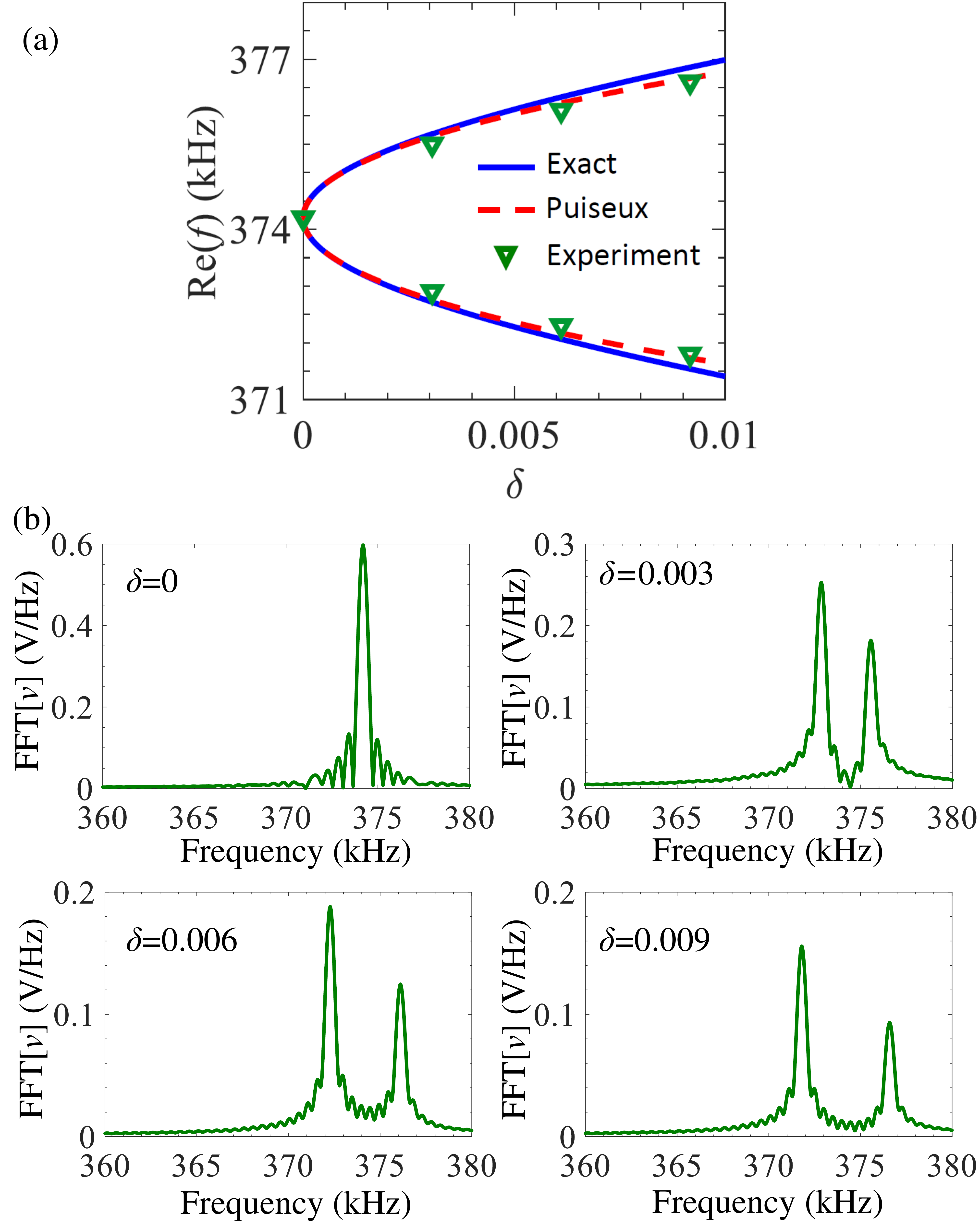}
\par\end{centering}
\caption{\label{fig:Sensitivity} (a) Proof of exceptional sensitivity. Experimental
and theoretical changes in the real part of the two resonance frequencies
$f_{p}$ due to a positive relative perturbation $\delta$ applied
to the capacitance $C_{1}$ of the time-varying capacitor (i.e., as
$(1+\delta)C_{1}$). The two real frequencies \textit{greatly depart}
from the EPD frequency around 374 kHz even for very small variation
of the capacitance following the fractional power expansion $\triangle f=f_{p}(\delta)-f_{es}\propto$$\sqrt{\delta}$.
The solid blue line, dashed red line and green symbols represent the
calculated-exact, Puiseux series approximation, and experimental resonance
frequencies of the LTP LC circuit shown in the Fig. \ref{fig:Real=000026Imag}(a).
The EPD frequency around 374 kHz corresponds to point \#1 (blue circle)
at the edge of the BZ in Fig. \ref{fig:exprmnt_results}(a). (b)
Fast Fourier transform of the measured capacitor voltage $v(t)$,
obtained at the EPD point and at three different perturbations of
the capacitor $C_{1}$. The four spectra correspond to the experimental
green triangles in Fig. (a) and they show the resonance frequency
shifts due to the capacitor's perturbation.}
\end{figure}

Sensitivity of a system's observable to a specific parameter is a
measure of how strongly a perturbation to that parameter changes the
observable quantity of that system. The sensitivity of a system operating
at an EPD is boosted due to the degeneracy of the system eigenmodes.
In the LTP system considered in this paper, a perturbation $\delta$
to a system parameter leads to a perturbed state transition matrix
$\underline{\boldsymbol{\Phi}}$ and thus to perturbed eigenvalues
$\lambda_{p}(\delta)$ with $p=1,2$. Therefore, the two degenerate
resonance frequencies occurring at the EPD change significantly due
to a small perturbation $\delta$, resulting in two distinct resonance
frequencies $f_{p}(\delta)$, with $p=1,2$, close to the EPD resonance
frequency. The two perturbed eigenvalues near an EPD are represented
using a convergent Puiseux series (also called fractional expansion
series) where the Puiseux series coefficients are calculated using
the explicit recursive formulas given in \citep{Welters2011Explicit}.
A first-order Puiseux approximation of $\lambda_{p}(\delta)$ is

\begin{equation}
\lambda_{p}(\delta)\approx\lambda_{e}+(-1)^{p}\alpha_{1}\sqrt{\delta}
\end{equation}
where $\alpha_{1}$ is a coefficient that is either purely real or
purely imaginary when losses can be neglected, and is given by

\begin{equation}
\alpha_{1}=\sqrt{-\frac{\frac{\mathrm{dD(0,\lambda_{e})}}{\mathrm{d}\delta}}{\frac{1}{2!}\frac{\mathrm{d^{2}D(0,\lambda_{e})}}{\mathrm{d}\lambda^{2}}}}.\label{eq:alpha_1}
\end{equation}

where $D(\delta,\lambda)=\left[\mathrm{det}[\boldsymbol{\underline{\Phi}}(\delta)-\lambda\underline{\mathbf{I}}]\right]$.
Since this is a second order polynomial in $\lambda$, the denominator
of (\ref{eq:alpha_1}) is equal to unity. The perturbed complex resonance
frequencies are approximately calculated as

\begin{equation}
f_{p}(\delta)\approx f_{es}\pm\mathrm{j}\frac{f_{\mathrm{m}}}{2\pi}(-1)^{p}\alpha_{1}\sqrt{\delta},\label{eq:omega_p}
\end{equation}
 where the $\pm$ signs correspond to the cases with EPD either at
the center ($f_{e0}=f_{\mathrm{m}}/2$ ) or edge ($f_{e0}=0$) of
the BZ, respectively (we recall that in this paper the fundamental
BZ is defined as $0\leq f/f_{\mathrm{m}}\leq1$ ). Equation (\ref{eq:omega_p})
is only valid for very small perturbations $\delta\ll1$ and it is
clear that for such a small perturbation the resonance frequencies
$f_{p}$ change dramatically from the degenerate resonance $f_{es}$
due to the square root function. In other words the EPD is responsible
for the square root dependence $\triangle f=f_{p}(\delta)-f_{es}\propto$$\sqrt{\delta}$.
Now, let us assume that the perturbation $\delta$ is applied to the
value $C_{1}$ of the time-varying capacitor, and the perturbed $C_{1}$
is expressed as $(1+\delta)C_{1}$. Considering an unperturbed LTP
LC resonator as shown in the subset of Fig. \ref{fig:Real=000026Imag}(a),
the system has a measured EPD resonance at a modulation frequency
$f_{\mathrm{m}}=62.8\,\mathrm{kHz}$ for the parameter values given
in section \ref{sec:Experimental-demonstration}. This EPD resonance
frequency is at $f_{e0}=0\,\mathrm{Hz}$, corresponding to point \#1
(blue circle) in the dispersion diagram shown in Fig. \ref{fig:exprmnt_results}(a),
and at all its Fourier harmonics $f_{es}=f_{e0}+sf_{\mathrm{m}}$.
By looking at the spectrum of the measured capacitor voltage we observe
that among the various harmonics of such EPD resonance, the frequency
of $\mathrm{Re}(f_{e6})=6f_{\mathrm{m}}=374.2\,\mathrm{kHz}$ has
a dominant energy component, hence it is the one discussed in the
following. The theoretical and experimental variations in the real
part of the two perturbed resonance frequencies due to a perturbation
$\delta\ll1$ in the the time-variant LC circuit are shown in Fig.
\ref{fig:Sensitivity}. Only the results for positive variations of
$\delta$ are shown here, hence the resonances move in the directions
where they are purely real (though the presence of small losses would
provide a small imaginary part in the resonance frequency). The solid
blue curve, dashed red curve, and green symbols denote the calculated-exact
(solutions of Eq. (\ref{eq:eigen value prob}) explained in Sec. \ref{sec:LPT_EPD_Theory}),
Puiseux series approximation, and the experimentally observed resonance
frequencies, respectively when varying $\delta$. 
The coefficient $\alpha_{1}$ in the Puiseux series is calculated
to be $\alpha_{1}=\mathrm{j}2.65$ which according to the fractional
expansion in Eq. (\ref{eq:omega_p}) it implies only change of the
real part of the resonance frequency for a positive perturbation $\delta>0$
while the imaginary part is constant. The three curves are in excellent
agreement for small perturbations, showing also the remarkable agreement
of the experimental results with the theoretical ones indicating that
this is a viable practical solution to make ultra-sensitive sensors.
The perturbation $\delta$ (the relative change in capacitance $C_{1}$)
is experimentally introduced through changing the \emph{positive}
voltage level of the pump voltage $v_{p}(t)$. In such a design, each
$5\,\mathrm{mV}$ change in the positive level of the pump voltage
will result in $1\%$ change of the $C_{1}$ capacitor value corresponding
to $\delta=0.01$ (see the Supplement Material). The
experimental results (green triangles) in Fig. \ref{fig:Sensitivity}(a)
represent the peaks of the fast Fourier transform (FFT) of the measured
voltage $v(t)$ of Node A with respect to the ground. The FFT in Fig.
\ref{fig:Sensitivity}(b) is taken over the time domain interval corresponding
to the ``ON'' state of the circuit, i.e., during 727$\,\mathrm{\mu s}$
which correspond to 80\% of 1.1 KHz, while $v_{reset}=0\,\mathrm{V}$(see
Supplemental Material). The results in Fig. \ref{fig:Sensitivity}
demonstrate theoretically and experimentally that for a small perturbation
$\delta$, the real part of the resonance frequency is significantly
changed and that it can be easily detected even in real noisy electronic
systems. Indeed, the spectral width of the measured peaks is small
enough to even distinguish the difference between $\delta=0$ and
$\delta=0.003$. We also increased the repetition frequency and observed
the linear-growth saturation of the system under the EPD regime, and
by doing the Fourier transform of the time domain signal in the four
cases over a time window of length 2 ms (which included the saturation
regime); this provided the same results provided in Fig. \ref{fig:Sensitivity}(a)
and (b). This latter test showed that the circuit can be operated
even including the saturation regime and it would still provide enhanced
sensitivity to perturbation. In conclusion, these experimental results
unequivocally demonstrate the exceptional sensitivity of the proposed
system operating at an EPD and the practicality of the LTP-resonator
circuit to conceive a new class of extremely sensitive sensors.

\subsection{Discussion}

A perturbation of the single resonator with an LTP
component (for instance a perturbation of the $C_{1}$ value of the
time-varying capacitance), perturbs the system away from the EPD.
In absence of loss, such a perturbation results in two real-valued,
shifted resonance frequencies$f_{p}(\delta)$. When losses are present
the EPD frequency $f_{es}$ is slightly complex (see Sec. \ref{sec:LPT_EPD_Theory})
and the two frequency shifts $\triangle f=f_{p}(\delta)-f_{es}$ are
approximately real valued. Instead, in a two-coupled resonator system
based on PT symmetry operating at an EPD, perturbing one of the capacitors,
for instance the one on the lossy (sensing) side, results in two complex
resonance frequencies (the value of $\alpha_{1}$ in that case would
be complex). The reasons of why the two shifted frequencies are complex
in a coupled-resonator circuit relies on the fact that the asymmetry
in the capacitances (due to the perturbation) disqualifies the system
as PT-symmetric and hence the resonance frequencies are no longer
real-valued. However, in the sensitive two-coupled PT-symmetric resonator
system presented in \citep{Chen2018Generalized,Sakhdari2018Ultrasensitive}
a (sensing) capacitance perturbation resulted in two real resonance
frequencies because in that system PT symmetry was maintained by manually
changing the capacitance on the gain side of the circuit (using a
varactor) to exactly balance the capacitance perturbation on the sensing
(lossy) side of the circuit.

Nevertheless, in many practical applications, a prior
knowledge of how much the sensing capacitance is perturbed is not
available since the amount of capacitance perturbation depends on
the physical (or chemical/biological) quantity to be measured, and
the perturbed frequencies would not be real. The reason of this striking
difference between the performance of a PT-symmetric coupled resonator
circuit and the proposed LTP single resonator circuit rely on the
fact that in the PT symmetry case the Puiseux series coefficient $\alpha_{1}$
for one varying capacitance is complex whereas in our case the Puiseux
series coefficient $\alpha_{1}$ for the varying capacitance is either
purely real or purely imaginary, depending on the EPD point considered
in the dispersion diagram in Fig. \ref{fig:Real=000026Imag}.

Another important consideration which shows a possible
advantage of our proposed sensing scheme is that the capacitance and
inductance values of electronic components are not precisely known,
i.e., commercially available components have prescribed tolerances
(e.g., 1\%, 2\%, 5\%). This uncertainty affects the exact occurrence
of the EPD in a PT-symmetric system and in turn it would affect its
sensitivity to perturbations. On the contrary, in our case, the tuning
of the modulation frequency would generate an EPD regardless of the
precise components values. Note that the exact frequency at which
the EPD occurs is not important in sensing applications since the
sensitivity is associated to the shift from such an EPD frequency,
and only the precise measurement if the shift is required.

Finally, it is important to note that electronic
and thermal noise in the proposed circuit did not compromise the capability
to experimentally verify the high sensitivity of the LTP single-resonator
circuit to a system perturbation, as clearly demonstrated in Fig.
\ref{fig:Sensitivity}. The resonance peaks in Fig. \ref{fig:Sensitivity}(b)
obtained from the measurement of the time domain voltage waveform
are very distinguishable from each other, and their spectral width
is much narrower than the frequency shift associated to even 0.3\%
variation of the perturbed capacitance. Therefore, this paper shows
the experimental proof of the existence of EPDs in a single resonator
circuit with a LTP modulation of a component, and also the high sensitivity
of the system's resonances to perturbations, regardless of the presence
of noise. Despite the topic of using EPDs to enhance sensitivity is
still subject to some debate, see for instance \citep{Langbein2018Noexceptional,Zhang2019Quantum,Lau2018Fundamental,Wiersig2020Prospects,Wiersig2020Robustness,Xiao2019Enhanced},
our experimental results clearly show that in some respect it is possible
to observe the special sensitivity provided by the square root behavior
$\sqrt{\delta}$ when $\left|\delta\right|\ll1$, proper of an EPD,
even in presence of realistic noise in the electronic system. Sensitivity
to perturbations could be further enhanced by understanding how the
parameter $\alpha_{1}$ could be increased by an improved design of
the system's components.

\section{Conclusion}

We have shown the first practical and experimental demonstration of
exceptional points of degeneracy (EPDs) directly induced via time
modulation of a component in a single resonator. This is in contrast
to EPDs realized in PT-symmetric systems that would require two coupled
resonators instead of one, and the precise knowledge of gain and loss
in the system as well as the values of L and C components to have
a high quality PT-symmetric resonator. Instead, in our proposed LTP-based
scheme we have shown that controlling the modulation frequency of
a component in a single resonator is a viable strategy to obtain EPDs
since varying a modulation frequency in a precise manner is common
practice in electronic systems. The occurrence of a second-order EPD
has been shown theoretically and experimentally in two ways: by reconstructing
the dispersion diagram of the system resonance frequencies, and by
observing the linear growth of the capacitor voltage. We have also
experimentally demonstrated how such a temporally induced EPD renders
a simple LC resonating system exceptionally sensitive to perturbations
of the system capacitance, and how the measurement of the shifted
frequencies is robust with respect to the presence of noise. Therefore,
the excellent agreement between measured and theoretical sensitivity
results demonstrate that the new scheme proposed in this paper is
a viable solution for enhancing sensitivity, paving the way to a new
class of ultra sensitive sensors that can be applied to a large variety
of problems where the occurrence of small quantity of substances shall
be detected.

It is important to observe that there are fundamental
differences between using the PT symmetry based circuit discussed
in \citep{Schindler2011Experimental,Sakhdari2018Ultrasensitive,Chen2018Generalized}
and the LTP circuit demonstrated in this paper, in detecting small
perturbations of a circuit element: (i) in the PT-symmetric based
circuit with two resonators, when the capacitance on one of the resonators
is varied the circuit is not PT-symmetric anymore and the two perturbed
resonance frequencies caused by the change of that capacitance are
always \textit{complex}-valued;
instead in our case, a perturbation of the capacitance leads to two
real-valued frequency shifts from the EPD one, and this may have very
important implications in sensing technology; (ii) our EPD is easy
to obtain by simply modifying the modulation frequency; (iii) the
capability to obtain EPDs is not sensitive to tolerances of realistic
components since we only need to tune the modulation frequency to
obtain an EPD (this is not true in PT symmetry systems where multiple
components needs to have precise values at the same time).
\begin{acknowledgments}
This material is based upon work supported by the National Science
Foundation under Award No. ECCS-1711975.
\end{acknowledgments}

\end{document}


\title{Supplemental Information:}
\title{Experimental Demonstration of Exceptional Points of Degeneracy in
Linear Time Periodic Systems and Exceptional Sensitivity}
\author{Hamidreza Kazemi, Mohamed Y. Nada, Alireza Nikzamir, Franco Maddaleno,
and Filippo Capolino}

\maketitle
We first define the time-varying capacitance based on the multiplier
concept in Fig. 3(a) in the main text and show that its definition
leads to an LC circuit described by differential equations consistent
to the previous theory shown in \citep{Kazemi2019Exceptional}, hence
leading to the occurrence of exceptional points of degeneracy (EPDs).
Then we discuss the physical implementation of the electronic linear
time periodic (LTP) LC resonator with a time varying capacitor and
reset signal.

The LTP LC resonator circuit shown in Fig. 3(c) in the main text is
composed of two subsystems: i) the sensing device which comprises
of a time variant LC resonator operating at the time-induced EPD \citep{Kazemi2019Exceptional},
and ii) a reset mechanism to stop the growing oscillation amplitude
and to set the initial voltage of the capacitor at the beginning of
each operation regime to avoid saturating the circuit. The schematic
and detailed working principle of each subsystem is explained in the
following sections.

\section{Time varying capacitance and fundamental dynamic equations}

The time varying capacitance is obtained by resorting to the multiplier
scheme in Fig. 3(a) in the main text. Without worrying for a moment
about the physical implementation and the reset scheme that are explained
in the following sections, we shall observe here that the voltage
$W(t)$ is given by the product $W(t)=v(t)v_{p}(t)/V_{0}$, where
the term $V_{0}$ is a constant coefficient of the multiplier that
is used to normalize its output voltage. Since in this paper the signal
$v_{p}(t)$ is a square wave with frequency $f_{m}$, it experiences
jumps (discontinuities), implying that also $W(t)$ is discontinuous.
The voltage applied to the capacitor $C_{0}$ is then given by $v_{c}(t)=v(t)-W(t)=v(t)(1-v_{p}(t)/V_{0})$.
This capacitor voltage cannot be discontinuous in time, hence $v_{c}(t)$
is a continuous function and indeed $v(t)$ carries the same discontinuities
as $W(t)$.

We define the capacitance of the synthesized time varying capacitor
used in this paper as $C(t)=q(t)/v(t)$, where $q(t)$ is the charge
on the capacitor. The same charge is also given by $q(t)=C_{0}v_{c}(t)$,
leading to the value of the time varying capacitance $C(t)=C_{0}v_{c}(t)/v(t)$.
Substituting for $v_{c}(t)$ in this latter equation leads to $C(t)=C_{0}(1-v_{p}(t)/V_{0})$
that is the synthesized piece wise time varying capacitance with period
$T_{m}$.

The two fundamental differential equations describing the LTP LC circuit
in Fig. 3(a) in the main text are $\mathrm{d}q/\mathrm{d}t=i_{c}(t)=-i(t)$
and $\mathrm{d}i/\mathrm{d}t=v(t)/L_{0}=q(t)/(L_{0}C(t))$, that in
matrix form are rewritten as

\begin{equation}
\frac{\mathrm{d}}{\mathrm{d}t}\left(\begin{array}{c}
q\\
i
\end{array}\right)=\left(\begin{array}{cc}
0 & -1\\
\dfrac{1}{L_{0}C(t)} & 0
\end{array}\right)\left(\begin{array}{c}
q\\
i
\end{array}\right).
\end{equation}

This first order differential equation is the same as that considered
in Eq. (1) in the main text and in \citep{Kazemi2019Exceptional}
and therefore it leads to the same dynamic current and voltage of
the LTP LC circuit studied in \citep{Kazemi2019Exceptional}, including
the occurrence of EPDs.

\section{Design and implementation of the time-varying capacitance}

The scheme of the proposed sensor (i.e., the time varying LC resonator)
is based on the design of a synthesized periodic time-varying capacitance
$C(t)$ that switches between two values $C_{1}$ and $C_{2}$ with
a modulation frequency $f_{m}$ where the parameters $C_{1}$, $C_{2}$
and $f_{m}$ need to be tunable. In order to make such a time-varying
capacitor, we use a regular time-invariant capacitor $C_{0}$ which
is connected in parallel with a multiplier $U_{1}$ (AD835 \citep{AD835})
as shown in Fig. \ref{fig:whole-circuit-sche}(a). In general terms,
the multiplier $U_{1}$ provides the time domain function

\begin{equation}
W(t)=\frac{(X_{1}-X_{2})(Y_{1}-Y_{2})}{V_{0}}+Z,
\end{equation}
that is connected to the lower terminal of the capacitor $C_{0}$
. Here $W$, $X_{1}$, $X_{2}$, $Y_{1}$, $Y_{2}$, and $Z$ represent
the voltages at the pins denoted with the same symbols of the multiplier
$U_{1}$ and $V_{0}=1.05\,\mathrm{V}$ is the voltage normalization
factor dictated by the multiplier and given in the data sheet \citep{AD835}.
In this circuit schematic, the $X_{2}$ and $Y_{2}$ pins are both
connected to the ground, the pump voltage $v_{p}(t)$ is applied to
the $Y_{1}$ pin and the $X_{1}$ pin is connected to Node $A$ as
shown in Fig. \ref{fig:whole-circuit-sche}(a). Hence, the output
function of the multiplier is simplified as

\begin{equation}
W(t)=\frac{X_{1}v_{p}(t)}{V_{0}}+Z.\label{eq:W}
\end{equation}
From the schematic and since the $X_{1}$ pin input current is zero
(it has high input impedance), we express the current flowing into
the capacitor $C_{0}$ as

\begin{eqnarray}
i_{c}(t) & = & C_{0}\frac{\mathrm{d}}{\mathrm{d}t}\left(X_{1}-W\right).\label{eq:Cap_curr}
\end{eqnarray}
The printed circuit board (PCB) layout and the actual assembled circuit
with resonator (capacitor $C_{0}$, inductor $L_{0}$, multiplier
chip $U_{1}$), chip with switches $U_{2}$, pump voltage $v_{p}(t)$
and reset $v_{reset}(t)$ connectors, and PCB traces are shown in
Fig. \ref{fig:whole-circuit-sche}(b) and (c). In the following we
describe the two operation regimes of the circuit and the principle
of the reset mechanism in detail.

\section{Regimes of operation}

The proposed circuit has two distinct phases (regimes) of operation
which are differentiated based on the status of the reset signal $v_{reset}$.
Reset signal $v_{reset}$ is a digital clock coming from an external
waveform generator that is used to stop the oscillation of the time-varying
LC resonator when it is high level voltage (logic Level 1), and allows
the resonator circuit to run when it is at low level voltage (logic
Level 0). These two different operation regimes are described in the
following.

\begin{figure*}
\begin{centering}
\includegraphics[width=6in]{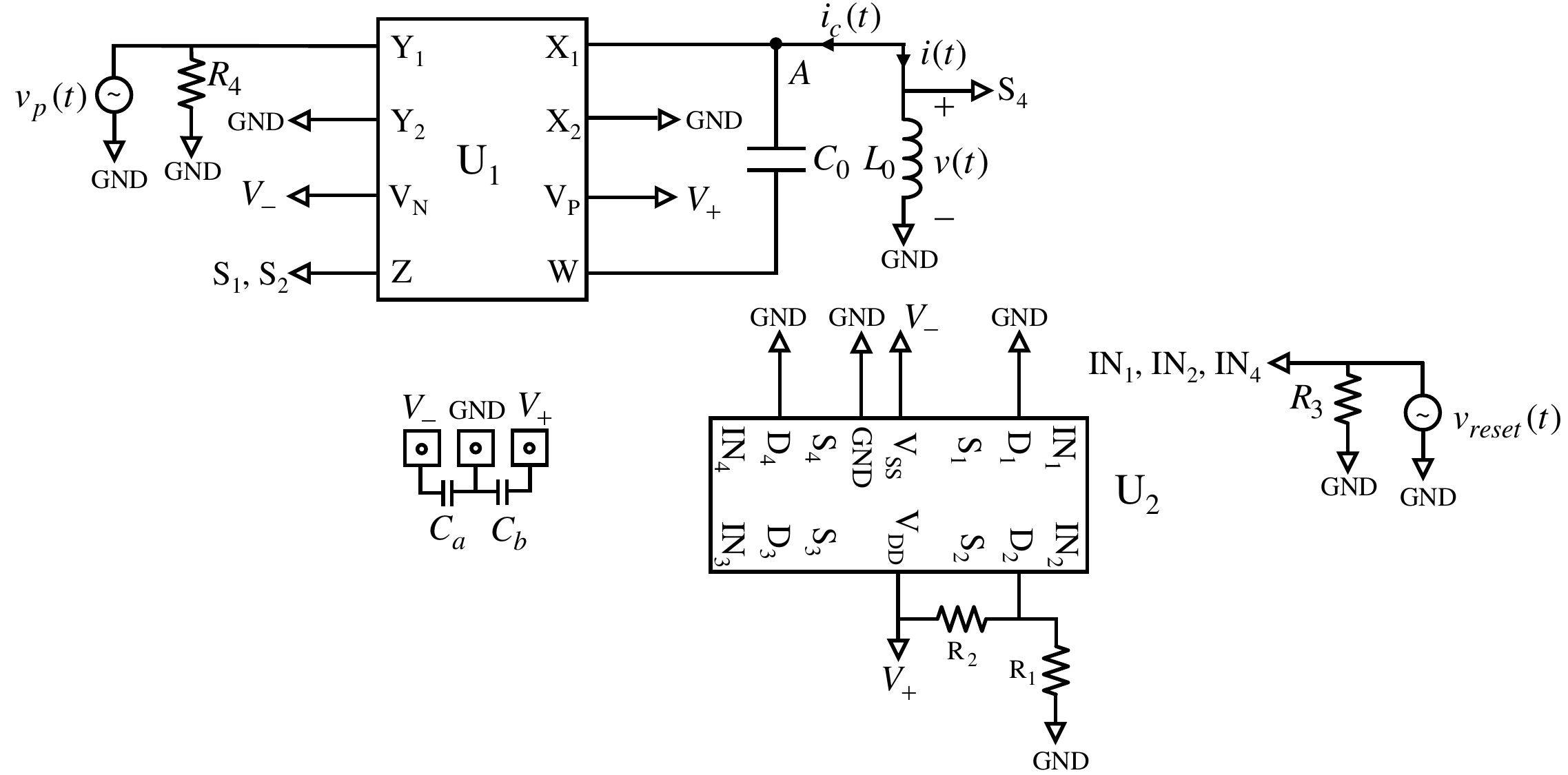}
\par\end{centering}
\begin{centering}
(a)
\par\end{centering}
\begin{centering}
\includegraphics[width=4in]{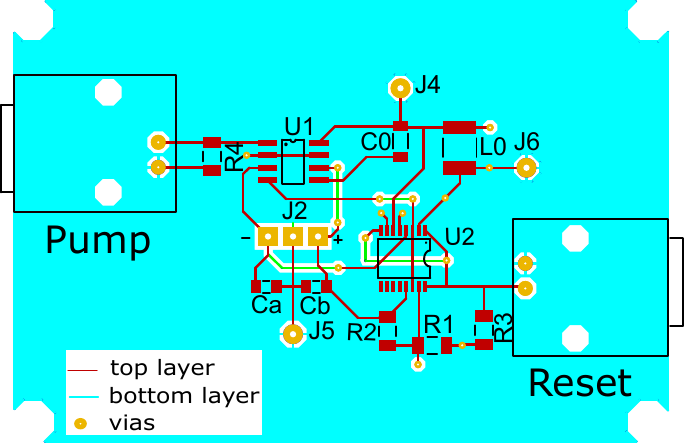}
\par\end{centering}
\begin{centering}
(b)
\par\end{centering}
\begin{centering}
\includegraphics[width=4in]{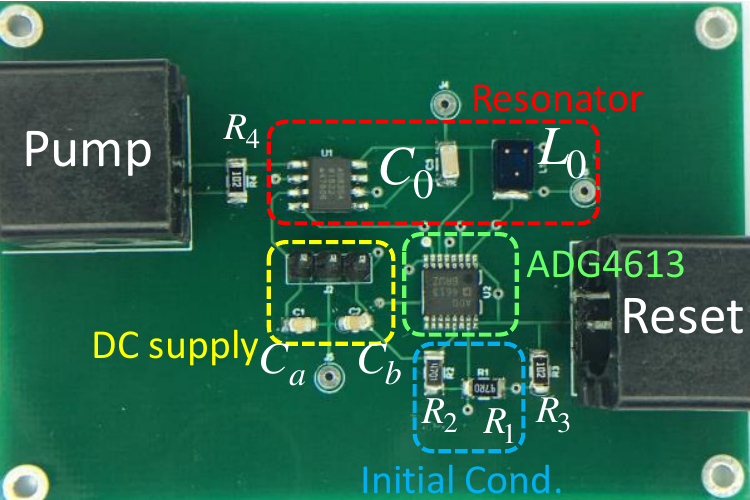}
\par\end{centering}
\begin{centering}
(c)
\par\end{centering}
\caption{\label{fig:whole-circuit-sche}(a) Schematic of the time-varying LC
resonator circuit based on the capacitor $C_{0}=10\,\mathrm{nF}$
(NP0 1206 \citep{CapNP0}), the inductor $L_{0}=33\,\mathrm{\mu H}$
(MSS7348-333ME \citep{Inductor}), the multiplier $U_{1}$ and the
reset mechanism circuit $U_{2}$. (b) PCB layout of the assembled
circuit where the top layer traces are in red, the ground plane is
in cyan, the bottom traces are in green, and the connecting vias are
in orange. In this design Via $J_{4}$ is a probe point for the capacitor
voltage, whereas Vias $J_{5}$ and $J_{6}$ are test points connected
to the ground plane and are used to connect the ground of the measurement
equipment to the ground of the circuit. All the ground nodes are connected
to each other using the bottom cyan layer. (c) Fabricated PCB and
assembled circuit with blocks highlighted.}
\end{figure*}

(i) \textit{LTP-LC operating regime} \textit{with EPD}: During this
regime, the reset signal is a logic low (logic Level 0) so that the
time varying LC resonator is allowed to have a free run, i.e., its
voltage is varying as in a time varying LC resonator with a given
initial voltage condition. During this regime, low level voltage $v_{reset}$
causes Switch $1$ to be ``on'', hence the pin $\mathrm{Z}$ of
the multiplier $U_{1}$ shown in Fig. \ref{fig:whole-circuit-sche}(a),
is connected to the ground (see section \ref{sec:Reset-mechanism})
while Switch $4$ is off, therefore, the input $X_{1}=v(t)$. In this
case, by using Eq. (\ref{eq:W}), the multiplier output is $W(t)=v(t)v_{p}(t)/V_{0}$.
Therefore the voltage difference at the two ends of the capacitor
$C_{0}$ is $v_{c}(t)=v(t)-v(t)v_{p}(t)/V_{0}$ and the current flowing
into the capacitor is

\begin{eqnarray}
i_{c}(t) & = & C_{0}\frac{\mathrm{d}}{\mathrm{d}t}\left(v(t)-v(t)v_{p}(t)/V_{0}\right),
\end{eqnarray}
which can be also interpreted as flowing into a synthesized time varying
capacitor with applied voltage $v(t)$. Considering a two level piece-wise
constant pump voltage $v_{p}(t)$, where $\mathrm{d}v_{p}(t)/\mathrm{d}t=0$
for $0<t<0.5T_{m}$ and $0.5T_{m}<t<T_{m}$, the current flowing into
the capacitance is given by 

\begin{equation}
i_{c}(t)=C(t)\frac{\mathrm{d}v}{\mathrm{d}t}=C_{0}\left(1-v_{p}(t)/V_{0}\right)\frac{\mathrm{d}v}{\mathrm{d}t}.\label{eq:Appx_ic}
\end{equation}

It is clear form Eq. (\ref{eq:Appx_ic}), that the capacitance seen
from Node $A$ towards the ground is time-variant and its shape is
dictated by the pump voltage. Considering the pump voltage as

\begin{eqnarray}
v_{p}(t) & = & \begin{cases}
0.525\,\mathrm{V} & 0<t<0.5T_{m}\\
-0.525\,\mathrm{V} & 0.5T_{m}<t<T_{m}
\end{cases},\label{eq:vp}
\end{eqnarray}
and recalling that $V_{0}=1.05\,\mathrm{V}$, the time varying capacitance
from Eq. (\ref{eq:Appx_ic}) is

\begin{eqnarray}
C(t) & = & \begin{cases}
0.5C_{0} & 0<t<0.5T_{m}\\
1.5C_{0} & 0.5T_{m}<t<T_{m}
\end{cases}.\label{eq:c(t)}
\end{eqnarray}

We observe from Eqs. (\ref{eq:Appx_ic})-(\ref{eq:c(t)}), that a
change of $5\,\mathrm{mV}$ in the high level of the pump voltage
$v_{p}(t)$ will result in approximately a relative change of $1\%$
in the capacitance value $C_{1}$, i.e.,

\begin{eqnarray}
\delta & \triangleq & \frac{C_{1,perturbed}-C_{1}}{C_{1}}\\
 & = & \frac{(1-0.520/1.05)C_{0}-0.5C_{0}}{0.5C_{0}}\approx1\%
\end{eqnarray}

(ii) \textit{Reset regime}: During this regime, the reset signal is
a logic high (logic Level 1), therefore Switch $4$ is ``on'' and
$\mathrm{S}_{4}$ takes the value of $\mathrm{D}_{4}$ that is grounded.
As a result Node $A$ is connected to the ground and the operation
of the time-varying LC resonator is halted. Moreover, during this
regime, the capacitor $C_{0}$ is charged again with an initial voltage
to be ready for the next\textit{ LTP-LC operating regime}. The operation
of the reset mechanism circuit is explained in the following section.

\section{Reset and LTP-LC operating regime mechanisms\label{sec:Reset-mechanism}}

The reset mechanism in this circuit is employed to (i) prevent the
saturation of the system due to the growth of the capacitor voltage
with time and to (ii) set an initial voltage on the capacitor $C_{0}$
in order to start a new \textit{LTP-LC operating regime}. The reset
mechanism is implemented using the ADG4613 chip which contains four
independent single pole/single throw (SPST) switches \citep{ADG4613},
denoted by $\mathrm{U}_{2}$ in the Supplementary Fig. \ref{fig:whole-circuit-sche}(a).
This figure shows the four switches where $\mathrm{IN}$, $\mathrm{D}$,
and $\mathrm{S}$ represent the control input, input, and output of
the switches, respectively. Two switches (1 and 3) are turned on with
logic Level 1 (high level voltage of $v_{reset}$) on the appropriate
control input, implying that during that time $\mathrm{S}_{1}$ and
$\mathrm{S}_{3}$ take the value of terminals $\mathrm{D}_{1}=0$V
and $\mathrm{D}_{3}$, respectively, and they are open when they are
``off''. Switch 3 is not used in our physical implementation. The
logic is inverted on the other two switches (2 and 4), i.e., the switches
turn on with logic Level 0 (low level voltage of $v_{reset}$) on
their control input implying that during that time $\mathrm{S}_{2}$
and $\mathrm{S}_{4}$ take the value imposed at terminals $\mathrm{D}_{2}=V_{+}R_{1}/(R_{1}+R_{2})$,
and $\mathrm{D}_{4}=0\,\mathrm{V}$, respectively. In the reset mechanism
we use only Switches 1, 2, and 4 where the control input of these
switches is connected to a digital clock signal $v_{reset}$ as indicated
in Supplementary Fig. \ref{fig:whole-circuit-sche}(a). The reset
signal is a digital clock with $20\%$ duty cycle, i.e., $v_{reset}=2\,\mathrm{V}$
for 20\% of its period (logic Level 1) and $v_{reset}=0\,\mathrm{V}$
otherwise (logic Level 0). Therefore, the LTP LC resonator is showing
an oscillating output for 80\% of the reset signal period. In summary,
during the 80\% of the $v_{reset}(t)$ period the circuit has $\mathrm{S}_{1}=0\,\mathrm{V}$,
while the $\mathrm{S}_{2}$ and $\mathrm{S}_{4}$ are open circuits,
implying that $Z=0\,\mathrm{V}$ and $X_{1}=v(t)$, whereas during
the remaining 20\% of the reset signal period, the circuit has $\mathrm{S}_{2}=V_{+}R_{1}/(R_{1}+R_{2})$
and $\mathrm{S}_{4}=0\,\mathrm{V}$ while $\mathrm{S}_{1}$ is left
open, implying that $Z=\mathrm{S}_{2}$ and $X_{1}=0\,\mathrm{V}$.

\subsection{Halting the oscillation}

As described, one of the benefits of the reset mechanism circuit is
to stop the time-varying LC resonator operating at an EPD before it
reaches saturation. For this purpose, during the \textit{reset regime},
Switch $4$ is used to short Node $A$ to the ground when the reset
signal $v_{reset}$ is at logic Level 1. Indeed the reset signal logic
Level 1 is connected to the input control $\mathrm{IN_{4}}$ which
turns on Switch $4$ connecting $\mathrm{S}_{4}$ to the ground. Since
Node $A$ is connected to $\mathrm{S}_{4}$, the voltage across the
inductor $L_{0}$ is set to zero preventing the oscillation .

\begin{figure}
\begin{centering}
\includegraphics[width=2.5in]{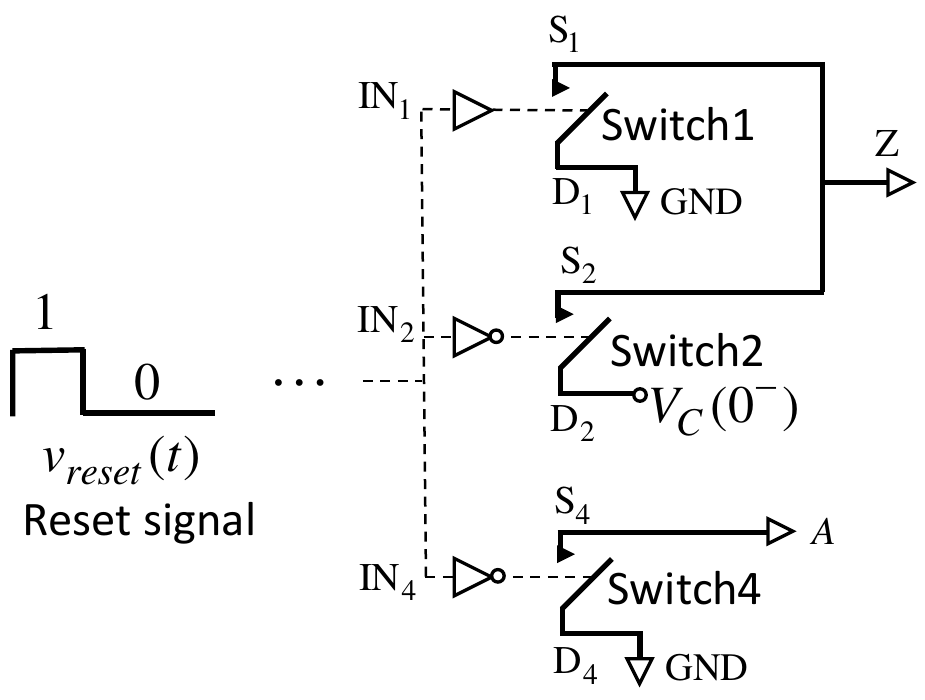}
\par\end{centering}
\caption{\label{fig:reset_IC}The three switches used for establishing the
two regimes of operation: the LTP LC resonator regime with EPD and
the reset mechanism that at the same time is also charging the capacitor
$C_{0}$ with an initial voltage to be used to start the next LTP
LC resonator regime. These switches are Switch 1, 2, and 4 of $\mathrm{U}_{2}$
shown in Fig. \ref{fig:whole-circuit-sche} and their control inputs
$\mathrm{IN_{1}}$, $\mathrm{IN_{2}}$, and $\mathrm{IN_{4}}$ are
connected to the reset signal $v_{reset}(t)$. The outputs of Switch
1 and 2, $\mathrm{S}_{1}$ and $\mathrm{S_{2}}$, are both connected
to the pin $Z$ of the $\mathrm{U}_{1}$, and $\mathrm{S_{2}}$ is
used to provide the initial voltage of the capacitor at the beginning
of each operating LTP LC regime. The output $\mathrm{S_{4}}$ of Switch
4 is connected to Node $A$ and is used to set the short the inductor
when the reset signal $v_{reset}(t)$ is at high voltage (logic Level
1).}
\end{figure}

\subsection{Setting the initial voltage condition on the capacitor $C_{0}$}

During each \textit{reset regime} just described, Node $A$ is grounded
therefore the voltage oscillation is halted, and the inductor current
decays with a short transient. In order to start a new \textit{LTP
LC operating regime} after each \textit{reset regime}, it is necessary
to provide the capacitor with the initial voltage. This is done by
using Switch 2 of the ADG4613 shown in the Supplementary Figure \ref{fig:reset_IC}
with its control input $\mathrm{IN_{2}}$ connected to $v_{reset}$.
Since $\mathrm{S}_{2}$ is connected to the $Z$ pin, during this
\textit{reset regime} time we have $Z=V_{+}R_{1}/(R_{1}+R_{2})$,
and $Z$ represents the summing input of the voltage multiplier as
shown in Eq. (2). Furthermore since Node $A$ is grounded, we have
$X_{1}=0\,\mathrm{V}$, resulting in $W=Z=V_{+}R_{1}/(R_{1}+R_{2})$,
i.e., the capacitor is inversely charged. An initial capacitor voltage
$V_{C}(0^{-})=-V_{+}R_{1}/(R_{1}+R_{2})=-50\,\mathrm{mV}$ is produced
by the voltage divider using $R_{1}=47\,\mathrm{\Omega}$ and $R_{2}=4.7\,\mathrm{k}\mathrm{\Omega}$
and by the positive DC supply $V_{+}=5\,\mathrm{V}$ shown in the
Supplementary Fig. \ref{fig:whole-circuit-sche}(a).